\documentstyle[preprint,aps,epsf,prc]{revtex}

\begin{document}
\title{P-parity of charmed particles from associative photoproduction of $D$ and $D^*$-mesons}
\author{Michail P. Rekalo \footnote{ Permanent address:
\it National Science Center KFTI, 310108 Kharkov, Ukraine}
}
\address{Middle East Technical University, 
Physics Department, Ankara 06531, Turkey}
\author{Egle Tomasi-Gustafsson}
\address{\it DAPNIA/SPhN, CEA/Saclay, 91191 Gif-sur-Yvette Cedex, 
France}
\date{\today}

\maketitle
\begin{abstract}
We suggest to measure the triple polarization correlations in the exclusive associative charm particle photoproduction, $\vec\gamma+\vec p \to \vec\Lambda_c^++\overline{D^0}$ with linearly polarized photons, as a method to determine the P-parity of the charmed D-meson. The dependence of these correlations on the parity $P(N\Lambda_cD)$ can be predicted in model independent way. The $t$-dependence of the differential cross section for vector meson photoproduction, $\gamma+ p \to \Lambda_c^++\overline{D^{*0}}$, in a model based on $D-$exchange, is also sensitive to $P(N\Lambda_cD)$.
\end{abstract}
\section{Introduction}
The P-parity is a fundamental property of any hadron, therefore it has to be experimentally determined. Generally  two experimental methods 
are used to determine the P-parity of elementary particles. 

For the meson and baryon resonances, which are unstable with respect to the strong or electromagnetic interactions, the P-parity can be determined through the selection rules with respect to angular momentum and P-parity, because  
they are conserved in such decays.

This method can not be applied to the stable or quasi-stable particles, which decay through weak interaction, where P-parity is not conserved. In this case the P-parity can be experimentally determined through specific scattering or production processes, induced by the strong or electromagnetic interaction \cite{Ga66}. A typical example is given by the K-meson \cite{Bi58,Bo59,Pa99}. First of all, due to the conservation of strangeness in the strong and electromagnetic interactions, the absolute value of the P-parity of the K-meson does not have a physical meaning, as in the case of $\gamma$, $\pi^0$, $\eta$, $\rho^0$..(which are characterized by zero values of the electric and baryonic charges, of strangeness and charm). Only the relative P-parity for the K-meson is significative, as in the case of the charged pion: we refer to the parity $P(\pi^{\pm})$ in a 'reference' frame, where the P-parities of proton and neutron are the same. Similarly for the K-meson: its strangeness is non zero, and, due to the conservation of strangeness in such process as $\gamma+N\to \Lambda+K$ or $p+p\to \Lambda+K^++p$  the K-meson P-parity is defined with respect to the $N\Lambda$-system.

In the lightest charmed hadron sector, all the decays of  the baryon $\Lambda_c^+(2285)$ and the meson $D(1870)$, through numerous channels due to the weak interaction, with violation of P-invariance, can not be used to determine their P-parity. On the contrary the P-parity of the vector charmed $D^*$-meson can be determined through its strong, $D^*\to D+\pi$ or electromagnetic $ D^*\to D+\gamma$-decays. More exactly, these decays determine the relative P-parity $P(D^*)P(D)$, and it is necessary, in any case, to know the P-parity of the lightest $D$-meson. The same is correct for the $\Sigma_c^+$-hyperon, with the strong decay $\Sigma_c \to \Lambda_c+\pi$ or with the electromagnetic decay $\Sigma_c^+\to \Lambda_c^++\gamma$ (for positive hyperons). Therefore, the $P(N\Lambda_cD)$-parity has to be experimentally determined, in order to have the P-parity 'reference' frame in the charmed particle sector.

The quark model gives clear predictions for the P-parity of charmed particles. For example, in case of $D^-=d\overline{c}$, for the  S-state of the quark-antiquark system, applying the Berestezky theorem \cite{Be72} about the negative internal P-parity of any fermion-antifermion system, one derives that the lightest D-meson must be a pseudoscalar particle. But one additional assumption has to be done: the P-parity of all quarks, with different quantum numbers is the same. Such assumption constrains all hadronic P-parities, and it is valid in the quark model.

The purpose of this paper is to suggest an experimental method to determine the P-parity of the lightest charmed particles, independently from the quark model.
We show here that the measurement of polarization phenomena in the simplest reaction of associative photoproduction of charmed particles, $\gamma+p\to \Lambda_c^++\overline{D^0}$, allows one to determine the relative P-parity $P(N\Lambda_cD)$ (but not the absolute D-meson P-parity). The charm and the P-parity are conserved in photoproduction processes.

\section{Polarization observables in $\gamma+p\to \Lambda_c^++\overline{D^0}$  }
The following discussion is based on the general symmetry properties of the electromagnetic interaction of charmed particles, without additional dynamical assumptions about the reaction mechanism for $\gamma+p\to \Lambda_c^++\overline{D^0}$.

It is well established \cite{Ch57}, that the spin structure of the matrix element for the process $\gamma+N\to {\cal B}_c+{\cal P}$, where ${\cal B}_c$ is a baryon with spin 1/2 and ${\cal P}$ is a meson with spin 0 is determined by a set of four independent amplitudes, which are complex functions of two independent kinematical variables, as the total energy $s$ and $\cos\theta$, where $\theta$ is the angle of the emitted baryon.

The spin structure of the  matrix element, in the CMS of the considered reaction,  can be parametrized in the following general form:
$$
{\cal M}^{(\pm)}=\chi^{\dagger}_2{\cal F}^{(\pm)} {\chi}_1,
$$
\begin{equation}
{\cal F}^{(+)}=\vec\sigma\cdot\hat{\vec q} {\cal F}^{(-)},~{\cal F}^{(-)}=\vec\sigma\cdot\vec e f_1-i\vec\sigma\cdot\hat{\vec q}\vec\sigma\cdot\hat{\vec k}\times \vec e f_2+ \vec\sigma\cdot\hat{\vec q}
\vec e\cdot\hat{\vec q}f_3+\vec\sigma\cdot\hat{\vec k}\vec e\cdot\hat{\vec q}f_4,
\label{eq:mat}
\end{equation}
where $\vec e$ is the real photon polarization vector, $\chi_1$ and $\chi_2$  are the
two-component spinors of the initial nucleon and the final baryon, 
$\hat{\vec k}$ and $\hat{\vec q}$ are the unit vectors along the three momenta of the $\gamma$ and the $D-$meson, respectively, in the reaction CMS. The upper indexes in ${\cal F}^{(\pm)}$ correspond to $P(N\Lambda_cD)=\pm 1$.

Generally, the polarization observables for the considered reaction are different for different values of  $P(N\Lambda_cD)$ and they can be calculated only in the framework of a definite dynamical model for the considered reaction
\cite{Re1,Re2,Re3}. However, in definite kinematical conditions it is possible to predict the exact value of polarization phenomena in model independent way. For example, the asymmetry of D-meson photoproduction, in collinear kinematics, induced by the collision of circularly polarized photons with a longitudinally 
polarized proton target is equal to $+1$, for any incident photon energy. This is a model independent result, which follows only from the helicity conservation in collinear kinematics, i.e. for $\theta=0^0$ or $\theta=\pi$. However, this result does not depend on $P(N\Lambda_cD)$ and it can not help to determine the parity of the charmed particles involved.

One has to find another polarization observable, which is sensitive to $P(N\Lambda_cD)$ on one side, and it is model independent, on the other side. The spin structure of the collinear matrix element for the process $\gamma+p\to \Lambda_c^++\overline{D^0}$ depends on $P(N\Lambda_cD)$:
\begin{eqnarray}
{\cal F}^{(-)}_{col}&=&\vec\sigma\cdot\hat{\vec e} f^{(-)}_{col},\nonumber \\
{\cal F}^{(+)}_{col}&=&\vec\sigma\cdot\vec e\times\hat{\vec k} f^{(+)}_{col}, 
\label{eq:mppf} 
\end{eqnarray}
where $f^{(\pm)}_{col}$ is the collinear amplitude for 
$\gamma+p\to \Lambda_c^++\overline{D^0}$ for $P(N\Lambda_cD)=\pm 1$.

Due to the presence of a single allowed amplitude in collinear kinematics, all polarization observables have definite numerical values, which are independent on the model chosen for $f^{(\pm)}_{col}$.

Let us consider the most general case, the dependence of the $\Lambda_c$ polarization on the polarization of the colliding particles. Using expressions (\ref{eq:mat}) for different $P(N\Lambda_cD)$, one can find the following formulas for the triple 
polarization  correlations in $\vec\gamma+\vec p\to \vec\Lambda_c^++\overline{D^0}$:
$$-(\vec e\cdot\vec e)(\vec P_1\cdot\vec P_2)+2(\vec e\cdot\vec P_1)(\vec e\cdot\vec P_2), \mbox{ ~if~ }  P(N\Lambda_cD)=-1,$$
\begin{equation}
(\vec e\cdot\vec e)[(\vec P_1\cdot\vec P_2)-2(\hat{\vec k}\cdot\vec P_1)(\hat{\vec k}\cdot\vec P_2)]-
2(\vec e\cdot\vec P_1)(\vec e\cdot\vec P_2), \mbox{~if~} P(N\Lambda_cD)=+ 1,
\label{eq:par} 
\end{equation}
where $\vec P_1$ and $\vec P_2$ are the polarization vectors for the initial and final baryons.

One can see from (\ref{eq:par}), that only the linear photon polarization affects the triple polarization correlations in $\vec\gamma+\vec p\to \vec\Lambda_c^++\overline{D^0}$, due to the P-invariance of the electromagnetic interaction of charmed particles. For further development, let us define the coordinate system for the considered collinear kinematics with  the $z-$axis  along $\hat{\vec k}$ and the $x-$ axis  along the vector ${\vec e}$ of the photon linear polarization. The correlations  (\ref{eq:par}) can be written in such system as:
$$-(\vec P_1\cdot\vec P_2)+2 P_{1x} P_{2x}= P_{1x} P_{2x}-P_{1y} P_{2y} -P_{1z} P_{2z}, \mbox{ ~if~ }  P(N\Lambda_cD)=- 1,$$
\begin{equation}
(\vec P_1\cdot\vec P_2)-2 P_{1z} P_{2z}-2P_{1x} P_{2x}=-P_{1x} P_{2x}+P_{1y} P_{2y} -P_{1z} P_{2z}, \mbox{~if~} P(N\Lambda_cD)=+ 1.
\label{eq:mpp2} 
\end{equation}

From (\ref{eq:mpp2}) one can find a connection between the components of the vectors $\vec P_1$ and $\vec P_2$ for the different $P(N\Lambda_cD)$, assuming,  for simplicity, that initially one has 100$\%$ linearly polarized photons:
$$P_{2x}= +P_{1x},~ P_{2y}=-P_{1y}, P_{2z}=- P_{1z}, \mbox{ ~if~ } 
P(N\Lambda_cD)=- 1,$$
\begin{equation}
P_{2x}= -P_{1x},~ P_{2y}=+P_{1y}, P_{2z}=- P_{1z}, \mbox{~if~} P(N\Lambda_cD)=+ 1,
\label{eq:mpp3} 
\end{equation}
One can see that both transversal components of the $\Lambda_c$-polarization are sensitive to $P(N\Lambda_cD)$, through the relative sign between $P_{2i}$ and $P_{1i}$:
\begin{eqnarray}
&P_{2x}=&P(N\Lambda_cD)P_{1x}, \nonumber \\
&P_{2y}=&P(N\Lambda_cD)P_{1y},
\label{eq:sol} 
\end{eqnarray}
whereas $P_{2z}=- P_{1z}$ for any value of $P(N\Lambda_cD)$.

Therefore, the relations ($\ref{eq:sol}$) allow one to determine, in model-independent way, the $D-$meson P-parity.

Let us summarize the main properties of the considered reaction, which are necessary for our consideration:

\begin{itemize}
\item The spin of the D-meson is equal to zero and the spin of the $\Lambda_c$-hyperon is equal to 1/2.
\item The P-parity is conserved in the $\gamma+p\to \Lambda_c^++\overline{D^0}$-process.  
\item The helicity is conserved in the collinear regime.
\end{itemize}

The suggested experiment, measuring the triple polarization correlations in 
$\vec \gamma+\vec p\to \vec \Lambda_c^++\overline{D^0}$, can be in principle realized by the Compass collaboration \cite{Compass}, which  has a polarized target and where linearly polarized photons can be obtained in muon-proton collisions, at small photon virtuality, by tagging the photon through the detection of the scattered  muon.

The $\Lambda_c$-polarization can be measured through the numerous weak decays of the  $\Lambda_c^+$-hyperon, for example $\Lambda_c^+\to \Lambda +e^+ +\nu_e$ \cite{PdG}, which is characterized by a large decay asymmetry. In other words, the $\Lambda_c$ is a self-analyzing particle. 

Eqs. (\ref{eq:sol}) show that only the relative sign of the transversal components of the polarization of the target proton and the produced  
$\Lambda_c^+$-hyperon is important for the determination of the 
$P(N\Lambda_cD)$-parity. Therefore such experiment does not need very large statistics, only well identified events. The energy of the photon beam has not to be necessarily monochromatic.

\section{Photoproduction of vector meson, $\gamma+ p \to \Lambda_c^++\overline{D^{*0}}$ }

Let us discuss the problem of the D-meson P-parity in connection to associative exclusive photoproduction of vector charmed mesons, $\gamma+\vec p \to \vec \Lambda_c^++\overline{D^{*0}}$ - with a more complicated structure of the corresponding matrix element. In the general kinematical case the twelve possible scalar amplitudes are complex and non-vanishing functions  of two independent kinematical variables. Even in collinear kinematics there are three independent collinear amplitudes. So this problem can not be solved without additional dynamical assumptions.  Let us consider, as an example, the D-meson exchange in $\gamma+ p \to {\cal B}_c +D^*$, at any photon energy and any $D^*$-production angle \cite{Re4}. 

It is possible to prove that the D-exchange mechanism has definite properties, which can be experimentally verified:
\begin{itemize}
\item The angular distribution of the decay products of the $D^*$-meson ($D^*\to D+\pi$), emitted in the collision of unpolarized particles,  $\gamma+p\to \Lambda_c^++D^*$, follows a $\sin^2\omega$-dependence, where $\omega$-is the angle between the photon momentum $\vec k$ and the three-momentum of $D$-in the rest system of $D^*$. This result depends on the relative P-parity of $D^*$ and $D$, but not on the 
$P(N\Lambda_cD)$-parity.

\item The beam asymmetry (due to linearly polarized photons) vanishes for any kinematical conditions, in $\gamma+p\to \Lambda_c^++\overline{D^{*0}}$, independently on  the $P(N\Lambda_cD)$-parity.
\item All the T-odd polarization observables vanish, because the D-exchange generates real amplitudes for $\gamma+p\to \Lambda_c^++\overline{D^{*0}}$.
\item The two possible asymmetries for the collisions of circularly polarized photons with a polarized proton target vanish - in any kinematical condition - due to the zero spin of the exchanged particle. Also this result is independent  on  $P(N\Lambda_cD)$. 
\end{itemize}

These facts can be considered as different experimental checks of the validity of $D$-meson exchange for $\gamma+p\to \Lambda_c^++\overline{D^{*0}}$. If it turns out that this approximation is valid, we can find some polarization effects which are sensitive to the $P(N\Lambda_cD)$-value. For example, the transversal components of the baryon polarizations $\vec P_1$ and $\vec P_2$ are related by the following condition:
\begin{equation}
P_{2x,y}=  P(N\Lambda_cD)P_{1x,y},
\label{eq:mds} 
\end{equation}
which is valid also in case of unpolarized photon beam, for any $D^*$-meson production angle. We must stress that  Eq. (\ref{eq:mds}) applies only in framework of the D-exchange model, but does not depend on many important characteristics of this model, such as the value of the coupling constants $g_{D^*D\gamma}$ and $g_{N\Lambda_cD}$ (in the vertices of the considered diagram) and on the phenomenological form factors, $F(t)$, which are necessary to insure the correct behavior of the differential cross section $d\sigma/dt$ with respect to the momentum transfer squared. 

Whereas all these ingredients are very important for the prediction of the absolute value of the unpolarized cross section, the shape of the $t-$dependence of the differential cross section, at small $|t|$ (for forward $D^*$-production) and at large photon energy is sensitive to  $P(N\Lambda_cD)$. This can be shown from the expression of the differential cross section, which can be written as:
$$\displaystyle\frac{d\sigma}{dt}(\gamma p\to  \Lambda_c^+ \overline{D^{*0}})={\cal N}(s)\displaystyle\frac{g_{D^*D\gamma}^2}{4\pi}
\displaystyle\frac{g_{N\Lambda_cD}^2}
{4\pi}F^2(t) \left (\displaystyle\frac{t-M^2}{t-m^2}\right )^2\left [-t+(M_N\pm M_{\Lambda})^2\right ].$$
Here ${\cal N}(s)$ is a known normalization factor, which depends on the photon energy, $M(m)$ is the $D^*$ $(D)$-mass, $M_N$ ($M_{\Lambda}$) is the nucleon (hyperon) mass and the signs $(\pm)$ correspond to $P(N\Lambda_cD)=\pm 1$.
 
The possible effect of different P-parities is shown in Fig. 1, at $E_{\gamma}=$ 40 GeV. The $t$-dependence of the differential cross section, normalized to its  maximum value (for $t=t_{max}=-0.23$ GeV$^2$), is represented as a solid (dashed) line for positive (negative) $P(N\Lambda_cD)$ . The different $t-$ behavior is due to the specific factor $\left [ -t+(M_N\pm M_{\Lambda})^2\right ]$, which is sensitive to the discussed P-parity. For simplicity we take the same (dipole) form factor $F(t)$ for both parities:
$$F(t)=\left [1-\displaystyle\frac {t-m^2}{m_0^2}\right ]^{-2},$$
where $m_0$=1.5 GeV is a cutoff parameter. These ratios do not depend on the values of the coupling constants $g_{D^*D\gamma}$ and $g_{N\Lambda_cD}$ and have a weak dependence on the form factor $F(t)$.

This prediction is valid only in the framework of the $D-$exchange model.

\section{Conclusions}
The triple polarization correlations in 
$\vec \gamma+\vec p\to \vec \Lambda_c^++\overline{D^0}$ in collinear kinematics and at any photon energy above the reaction threshold - with linearly polarized photons - can be suggested as a model independent method for the determination of the D-meson P-parity, or more exactly, the relative P-parity of the $N\Lambda_cD$-system. Note that in the standard quark model, this P-parity is negative. Another possible way to determine the D-meson P-parity, but model-dependent, is the study of the dependence of the $\Lambda_c^+$-polarization on the initial proton polarization in the framework of D-exchange for vector meson photoproduction $\gamma+\vec p \to \vec \Lambda_c^++\overline{D^{*0}}$, using an unpolarized photon beam. Note that the $t-$dependence of the differential cross section $d\sigma(\gamma p\to \Lambda_c^+ \overline{D^{*0}})$ is also characteristic of  the $P(N\Lambda_cD)$-parity, but the interpretation of the results depends on model assumptions, which are  necessary for the calculation of the differential cross section.

\begin{figure}
\mbox{\epsfysize=15.cm\leavevmode \epsffile{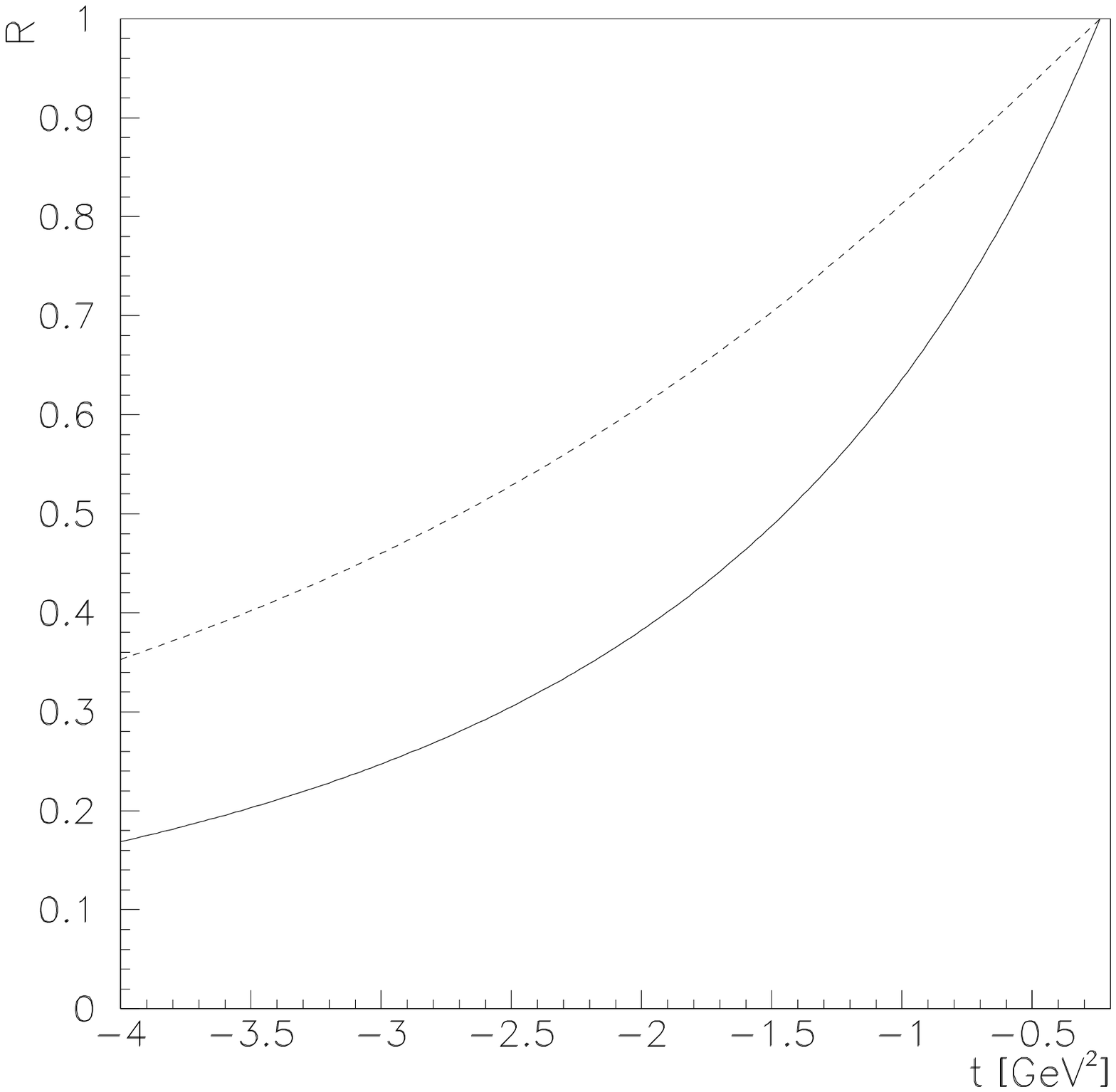}}
\vspace*{.2 truecm}
\caption{ $t-$dependence of the ratio 
$R=\displaystyle\frac{d\sigma}{dt}(\gamma p\to  \Lambda_c^+ \overline{D^{*0}})/
\displaystyle\frac{d\sigma_m}{dt}(\gamma p\to  \Lambda_c^+ \overline{D^{*0}})
$, where $\displaystyle\frac{d\sigma_m}{dt}(\gamma p\to  \Lambda_c^+ \overline{D^{*0}})$ is the  differential cross section for $t=t_{max}=-0.23$ GeV$^2$, for $P(N\Lambda_cD)=+1$ (solid line) and $P(N\Lambda_cD)=-1$ (dashed line), at $E_{\gamma}$=40 GeV.
}
\label{fig:fig1}
\end{figure}

\end{document}